\documentclass[sigconf]{acmart}

\AtBeginDocument{%
  }

\usepackage{amsmath,amsfonts}
\usepackage{algorithmic}
\usepackage{graphicx}
\usepackage{textcomp}
\usepackage{xcolor}
\usepackage{colortbl}
\usepackage{fontawesome}
\usepackage{etex}
\usepackage{booktabs}
\usepackage[utf8]{inputenc}
\usepackage[ruled,vlined]{algorithm2e}
\usepackage[T1]{fontenc}
\usepackage{microtype}
\usepackage{graphicx}
\usepackage{paralist}
\usepackage{tabularx}
\usepackage{soul}
\usepackage{balance}
\usepackage{multicol}
\usepackage{multirow}
\usepackage{pbox}
\usepackage{enumitem}
\usepackage{lscape}
\usepackage{pifont}
\usepackage{xspace}
\usepackage{url}
\usepackage{tikz}
\usepackage{float}
\usepackage[TABBOTCAP]{subfigure}
\usepackage{ragged2e}
\usepackage{fontawesome}
\usepackage[figuresright]{rotating}
\usepackage{bbding} 
\usepackage{adjustbox}
\usepackage{pifont}

\newcommand{\ie}{\emph{i.e.,}\xspace}
\newcommand{\eg}{\emph{e.g.,}\xspace}

\newcommand{\etal}{\emph{et~al.}\xspace}
\newcommand{\secref}[1]{Section~\ref{#1}\xspace}

\newboolean{showcomments}

\definecolor{lightgrey}{rgb}{0.925, 0.925, 0.925}
\sethlcolor{lightgrey}

\setboolean{showcomments}{true}

\ifthenelse{\boolean{showcomments}}
{\newcommand{\nb}[2]{
		\fbox{\bfseries\sffamily\scriptsize#1}
		{\sf\small$\blacktriangleright$\textit{#2}$\blacktriangleleft$}
	}
}
{\newcommand{\nb}[2]{}
}

\setcopyright{acmlicensed}
\copyrightyear{2018}
\acmYear{2018}
\acmDOI{XXXXXXX.XXXXXXX}

\acmConference[SE 2030]{International Workshop on Software Engineering in 2030}{November 2024}{Puerto Galinàs (Brazil)}
\acmISBN{978-1-4503-XXXX-X/18/06}

\begin{document}

\title{The Rise and Fall(?) of Software Engineering}

\author{Antonio Mastropaolo}
\orcid{0000-0002-7965-7712}
\affiliation{%
	\institution{Universit\`a della Svizzera Italiana}
	\country{Switzerland}
}
\email{antonio.mastropaolo@usi.ch}

\author{Camilo Escobar-Vel\'asquez}
\orcid{0000-0001-8414-9301}
\affiliation{%
  \institution{Universidad de los Andes}
  \country{Colombia}}
\email{ca.escobar2434@uniandes.edu.co}

\author{Mario Linares-V\'asquez}
\orcid{0000-0003-0161-2888}
\affiliation{%
	\institution{Universidad de los Andes}
	\country{Colombia}}
\email{m.linaresv@uniandes.edu.co}

\renewcommand{\shortauthors}{Mastropaolo et al.}

\begin{abstract}
	Over the last ten years, the realm of Artificial Intelligence (AI) has experienced an explosion of revolutionary breakthroughs, transforming what seemed like a far-off dream into a reality that is now deeply embedded in our everyday lives. AI's widespread impact is revolutionizing virtually all aspects of human life, and software engineering (SE) is no exception.

	As we explore this changing landscape, we are faced with questions about what the future holds for SE and how AI will reshape the roles, duties, and methodologies within the field. The introduction of these groundbreaking technologies highlights the inevitable shift towards a new paradigm, suggesting a future where AI's capabilities may redefine the boundaries of SE, potentially even more than human input.

	In this paper, we aim at outlining the key elements that, based on our expertise, are vital for the smooth integration of AI into SE, all while preserving the intrinsic human creativity that has been the driving force behind the field. First, we provide a brief description of SE and AI evolution. Afterward, we delve into the intricate interplay between AI-driven automation and human innovation, exploring how these two components can work together to advance SE practices to new methods and standards.
	
\end{abstract}



\received{5 April 2023}
\received[revised]{DD MMMM YYYY}
\received[accepted]{DD MMMM YYYY}


\maketitle


\section{Introduction}

With the exponential growth of AI capabilities, its integration into software engineering (SE) practices has revolutionized the way we approach development challenges, optimize and automate processes, and manage software engineering projects. From its beginning, AI has been looking to help humans on the execution of their daily life tasks, starting with automation of processes via robots, followed by the introduction of expert systems that could provide high quality results based on knowledge and  examples from human experts. Later on we saw the introduction of languages as LISP focused on making intelligent systems, the creation of chatbots, the introduction and extension of neural networks for the improvement of  learning capabilities of machines, and the definition of new models to enhance various domains including natural language processing, computer vision, robotics, and healthcare, among others.

Aforementioned improvements have helped different professionals in the automation of their daily tasks, and software engineering is not the exception; the processes and practices followed by developers, testers, managers and different roles of the software development life cycle are been supported on AI methods and tools.

Recent progress in AI-driven methods is crucially reshaping the realm of software engineering, leading to continuous modifications and conversations about the evolving role and input of humans in this domain. Developers now have tremendous aid to design and implement code; testers can rely on AI approaches to ``draft'' test cases; managers can automate the review process and automatically generate insights/visualizations using data extracted from the CI/CD processes. Then, what should be the role of software engineers in this new age? What has not been automated within the development process (yet) that still requires the interaction of humans? Are software engineers going to be unnecessary in the future? Is it the fall of software engineering?

This article presents a seminal discussion around the history and impact of AI on the development of software engineering as a field of study; in addition, we discuss the role of the software engineer in the current world. Therefore, \secref{sec:se_rise_fall} and \secref{sec:aihistory} present a brief history of Software Engineering and  AI; \secref{sec:ai4se} presents an outline of the current role of AI in SE, depicting the latest approaches generated in the 2020's; \secref{sec:discussion} presents a discussion on the newest AI technologies for SE, and the effects of the AI in the exercise of SE. Finally in \secref{sec:conclusion}  we delve into the future role of AI in SE and the upcoming challenges for software engineers.


\section{The Rise of Software Engineering}
\label{sec:se_rise_fall}

Software development was conceived as a sub-field of computer science, that in the early days of computing was focused mostly on the product (\ie the software) rather that in the process. However, the whole SE history has demonstrated that both (process and product) have been a main concern. A quick look to the latest version of the \textit{The Guide to the Software Engineering Body of Knowledge (SWEBOK)}~\cite{ieee2014} shows 15 knowledge areas that summarize key concepts of the SE discipline, including practices for each of the "steps" of the software development lifecycle as well as processes and transversal aspects (\eg management and economics).  It is worth noting that the term Software Engineering started to be widely used later in computing history (late 1960's), however,  there are records of earlier usage of the term; for instance, Margaret  Hamilton is probably the first person who came up with the idea of using "software engineering" as the name of a discipline different than computing  and hardware engineering \cite{CAMERON2018}.

The Software Engineering field was born as a natural reaction to the first documented "software crisis"\footnote{The "software crisis"  term was coined in the 1968 NATO SE Conference \cite{Naur1968}}, which refers to a period of time in which the creation of software systems was characterized by many difficulties such as low quality, unfinished projects or projects running over-budget and time, products that did not meet users' requirements, and projects with unmaintainable code. The 1960's software crisis motivated the creation of a NATO Science Committee (in 1967) and a Computer Science Study Group focused on the problems of software.  One of the recommendations of that group was to hold a working conference on Software Engineering, and this if probably the first time that the two words "software engineering" were mentioned as part of a working group document. The choice of those two words was made on purpose as a provocative action for the computing community, recognizing the need for (i) improving current practices of software development, and (ii) rethinking the process based on the well-established engineering disciplines (\eg civil and mechanical engineering).  Afterwards, the first SE conference was held in Garmish (Germany) in 1968  \cite{Naur1968}.

The 1968 NATO SE Conference \cite{Naur1968}  promoted the discussion around different topics such as (i) the role of SE in society; (ii) current practices on software design, production and services; (iii) problem areas, causes, and solutions; among other topics. It is worth noting that at that time, the established practices for software development were mostly for small systems. The increasing complexity of problems to be solved/tackled with software,  the rapid increase of computing power, and the lack of adequate  methods/techniques for creating large and complex software systems, led to several documented and well-known cases of software projects with the aforementioned difficulties.  Therefore, the next years after the NATO conference represent a tremendous effort in fighting the crisis by first identifying the causes and then mitigating the risks by using manufacture-based processes (\eg the waterfall method). 

In the ``post-NATO'' conference period there was a plethora of efforts devoted  (on the one side) to define and promote SE heavy/ bureaucratic processes and methods for increasing productivity and mitigating risks, but also (on the other side) to discuss the mistakes that have been done on software projects.  For instance, the waterfall method emerged as the \textit{de-facto} choice for building large systems, by relying on heavy steps of analysis and design that would provide developers with sufficient knowledge about the context, problem and the to-be system, before starting the coding; the heavy up-front analysis and design efforts would also assure a better planning of resources before the implementation. The details of the ``waterfall'' method were presented in the  \textit{Managing the development or large systems} paper written by Winston Royce in 1970 as a set of personal experiences after working for almost a decade in software projects for spacecraft mission planning \cite{Royce1987}.  Note that despite the waterfall process was used widely in a linear fashion, the paper also presented an iterative version of it.  As another remarkable example, later in 1981, Barry Bohem  started the discussion on  ``software economics''  \cite{Boehm1981} as a way for using economic analysis techniques in  software development with the purpose of having  better costs estimation and making better decisions.

On the side of the group of efforts discussing the causes, \textit{The mythical man-month}  book (1975)  \cite{Brooks1975}  by Fred Brooks, is as a set of essays on SE and management that discuss several causes of scheduling failures; famous thoughts there are the following:  ``measuring useful work in man-months is a myth''; ``adding  manpower to a late software project makes it later''; software construction involves essential and accidental tasks; and there is no silver bullet ``to make software costs drop as rapidly as software costs do''.

The writing of the \textit{Agile manifesto} in 2001 \cite{beck2001} represents a singularity in the SE history.  Different experiences, successful histories, but also failures, led a group of software engineers to share their thoughts and posit  a radical position on how we should be doing software. Since 1970, the SE discipline evolved with a plethora of methodologies and techniques for creating better software; linear, heavy,  iterative and incremental, lightweight processes, as well as widely adopted practices for each moment of the life cycle (\eg testing); supporting tools were also part of the SE book of knowledge. However, there were still evidences of  the "software crisis" even 30+ years after the 1968 NATO conference. The 1999 CHAOS report by the Standish Group \cite{Standish1999}, showed that in 1998 only 26$\%$ of the analyzed software of projects were successful.  Therefore the born of the software agility community  in 2001 was like a new rise of SE against the software crisis but with a new perspective that gives more value to (i) individuals and interactions over processes and tools; (ii) working software over comprehensive documentation; (iii)customer collaboration over contract negotiation; and (iv) responsive to change over following a plan. The four values and 12 principles of agile software (written in the agile manifesto) consolidate a history of about 50 years of the software development community. Since the mythical man-month thoughts to the radical propositions of the Extreme Programming methodology, the agile manifesto was like a new breath that triggered an interesting discussion around the nature of software development. Nowadays, the ``agile'' movement have even permeated other disciplines like management, product design, information  technology, among others.

In the ``post-agile manifesto'' times we have seen improvements on the process-perspective of software development but also on the product one. Concerning the latter, automated SE attracted a lot of attention by transferring techniques from other domains (\eg soft computing, bio-inspired computing, statistical learning, data analytics) and by having outstanding progress on formal verification methods. In addition, the availability of large sets of source coda via open-source projects and public repositories triggered a tremendous  interest towards learning from previous experiences documented on code/textual repositories (\eg issue trackers, social networks). By considering source code as text, software engineers were able to transfer, explore and explode techniques from disciplines like information retrieval, natural language processing, and machine learning (ML). Perhaps the first example on that line of research is the paper by Maletic and Marcus on the usage of information retrieval methods to support program comprenhension \cite{maletic2001}.  By considering source code as text, software engineers have enabled the automation of tasks such as software categorization \cite{Kawaguchi2004,Tian2009,McMillan2011}, bug triage \cite{Linares2012,Mani2019,Zhang2020}, defect prediction \cite{wang2016, kamei2016,Wan2020}, code completion \cite{Robbes2008, Raychev2014, Proksch2015}, automated patching/program repair \cite{kim2013,goues2019,Zhang2023}, documentation generation \cite{1McBurney2014,McBurney2016,Robillard2017}. 

After those initial steps on using information retrieval and shallow ML methods, the SE community started to explore different methods that led us to have new sub-fields such as Mining Software Repositories (MSR) and Natural Language-based Software Engineering (NLBSE), but also to include ML techniques as one of the mandatory tools in the software engineer toolbox. Seminal work on that line of thoughts are the publications on the  naturalness of software \cite{hindle2012}, deep learning (DL) repositories \cite{white2015}, and machine learning for big code \cite{allamanis2018}, which established the foundations for the current trend on using Large Language Models (LLMs) for SE. Nowadays, machine learning for software engineering,  software engineering for machine learning, and software engineering by and for humans, are new perspectives that define research and development avenues for the SE community  in the short, medium and large terms. We are perhaps in a new singularity in the SE history, promoted by the usage of LLMs and generative AI; therefore, before discussing the implications of artificial intelligence for the future of SE, we provide the reader (in the following sections) with a brief of the history of AI and its usage in SE.

\section{A brief history of Artificial Intelligence}
\label{sec:aihistory}

The history of artificial intelligence (AI) traces back to the mid-20th century, marked by groundbreaking developments and significant milestones that have shaped its evolution into the transformative technology we recognize today. Starting by the definition of fundamental terms as heuristic \cite{polya2004solve}, the analysis of the mathematical correspondence between nervous activity and binary machines\cite{mcculloch1943logical}, the study of computer-based assistance in different daily activities \cite{bush1945we}, the definition of a test to analyze intelligent behavior in computers \cite{turing2009computing} and the study of different math theories\cite{whitehead1997principia}, researchers founded the path towards analyzing and researching the cognitive capabilities a machine might have, like for example thinking or reasoning.

Based on these researchs, around 1956, John McCarthy, Marvin Minsky, Nathaniel Rochester, and Claude Shannon proposed the term "Artificial Intelligence" as main topic for the Dartmouth Conference, in which they proposed a two month summer research project focused on studying the current state of artifical intelligence and the aspects that might be considered within this topic: (i)~automatic computers, (ii)~how can a computer be programmed to use a language, (iii)~neuron bets based on the work by Pitts and McCulloch, (iv) ~theory of the size of a calculation, (v)~self-improvement, (vi)~abstractions, and (vii)~randomness and creativity \cite{dartmouth1956AI}.

Aftwerward, Frank Rosenblatt, a psychologist from Cornell University proposed a novel idea called ``perceptron'', that became a fundamental part of Neural Networks. The perceptron concept depicts the first computer that uses trial-and-error method to learn new information, specifically in its basic form: it learns the coefficient that represent the weights of the inputs for a linear equation built to perform classification. This capability is known as one of the main approaches of the artificial narrow intelligence (ANI). Based on the idea of the perceptron, Rosenblatt proposed a neural network model as a set of layers composed of perceptrons; these layers allowed machines  to perform harder tasks. In 1969, Marvin Minsky, participant of the Dartmouth Conference, and Seymour Papert published the book \textit{Perceptrons: an introduction to computational geometry}\cite{minsky2017perceptrons} in which they analyzed the mathematical implications behind two layer neural networks. After its publication, it was verified that their findings about limitations of the model were incorrect, but it opened the path for a technique called backpropagation, which is a method to adjust the weights in multilayer neural networks by compesating the error obtained during the learning process.

Around 1958, John McCarthy, participant of the Dartmouth Conference, invented the LISP language, that later  became the dominant programming language for AI. Using LISP, James Slagle created SAINT and present it in his PhD dissertation: \textit{A heuristic program that solves symbolic integration problems in freshman calculus}. SAINT was the first version of a machine that was able to support the decision process of humans by providing support via automation of cumbersome tasks. LISP and SAINT played a main role later on the history as the base for a set of machines called "expert systems".

Following this, in 1966, Joseph Weizenbaum developed ELIZA, an early natural language processing program capable of simulating conversation. ELIZA was created using MAD-SLIP an extension to fortran that followed several rules defined in LISP for processing and managing data. ELIZA used simple pattern matching and substitution techniques to answer queries, nevertheless, its main impact in the world of AI was a version programmed to ``simulate'' the dialogue of a psychotherapist, allowing ``expert systems'' not only to help researches to take decisions but also to retrieve information and process it.

After a few years of research and development in AI, the proposed approaches started to overcome the computation capabilities of the existing machines, leading the interest of the community towards the analysis and improvement of hardware prior to continuing the research in AI. Based on this decision, several funds started to prioritize the research of hardware improvement and stop funding research projects on AI; that time period (1974 to 1980) when the funds for AI were frozen was know as the AI Winter. This period was very short lived as new breakthroughs paved way for the second wave of AI.

In the 1980’s, AI was reignited by two aspects: an expansion of the algorithmic toolkit, and a boost of funds. Regarding the former aspect, the expansion started by the definition of "expert systems" that mimicked the decision making process of a human expert, taking into account a defined situation and the prior decision of an expert on a similar situation; these systems were defined to provide non-experts with advice based on experts knowledge. There are several examples of these systems, but Buchanan, Feigenbaum and Lederberg were the first to introduce "expert systems" in daily problems \cite{feigenbaum1970generality}; their work resulted in the creation of DENDRAL a system capable of assisting chemists with complex euclidean chemical structures. Based on this, the field of Machine Learning (ML) started growing by the definition of different algorithms that enhanced the type of support that could be provided by the creation of methematical models that supported classification, regression and reinforcement learning. Afterwards, John Hopfield and David Rumelhart popularized ``deep learning'' techniques \cite{hopfield1982neural,rumelhart1986learning} which reduced the human dependency by performing feature extraction as part of the learning process. As result of this DL techniques, new approaches were created enhancing another ANI fundamental task called reinforcement learning. Based on the learning experience and the definition of rewards and punishment, researchers of Stanford leaded by Hans Moravec created the first autonomous car that demonstrated the ability to navigate obstacles in its environment.

During the following 10 years, researchers continued working in ML and AI techniques, publishing and proposing new ideas, reducing the limitations to access expert systems and LISP machines, and focusing in demonstrations of machine learning, intelligent tutoring, case-based reasoning, multi-agent planning, scheduling, uncertain reasoning, data mining, natural language understanding and translation, vision, virtual reality, games, among others. One of the main milestones achieved during this time was the creation of Deep Blue, a chess program capable of beating the world chess champion, Garry Kasparov, marking a significant milestone in AI's ability to tackle complex problems. At the same time, other companies as NASA were using autonomous robotics systems to be deployed in the surface of Mars for its exploration. Additionally, Web crawlers and other AI-based information retrieval systems become primordial for the expansion of the World Wide Web \cite{cho2000evolution, chakrabarti1999focused}

These last systems marked a milestone in the development of the AI, since most of the current approaches and techniques rely mainly on the amount and quality of data available. For example, in 2011, IBM's Watson was able to defeat former "Jeopardy!" champions Brad Rutter and Ken Jennings in a televised trivia game, showcasing advancements in natural language processing and machine learning. Starting in this time, AI has been evolving continuously; in 2012 Convolutional Neural Networks (CNNs) demonstrated significant performance improvements in image recognition tasks, with AlexNet winning the ImageNet Large Scale Visual Recognition Challenge. Following by DeepMind's AlphaGo program that in 2016 defeats world champion Go player Lee Sedol, depicting an improvement on AI's ability to master complex games with vast decision spaces.

Finally in the late 2010's and early 2020's we have seen the continued advancements in various domains including natural language processing, computer vision, robotics, and healthcare, among others. Specifically for the SE field, the creation of LLMs and its extension to Generative AI techniques, that are used to support the software development process, via generation of code, tests and scripts via prompts and queries expressed in natural language.


\section{The Rise of Artificial Intelligence in Software Engineering (AI4SE)}
\label{sec:ai4se}
\vspace{0.2cm}

Artificial Intelligence (AI) has dramatically transformed the field of software engineering, moving from its initial applications \cite{le2011genprog,ali2009systematic,neil2003software} to current advanced techniques \cite{copilot,chatgpt} that assist developers and practitioners across a wide range of tasks related to SE \cite{watson2022systematic}.

The integration of AI into software engineering has evolved from basic methods, including computational/optimization search (\eg Search-Based Software Engineering - SBSE), probabilistic approaches, and Machine Learning (ML), to cutting-edge techniques underpinned by DNNs (Deep Neural Networks) having the ability to learn patterns from data by automatically extracting salient features for a given computational task as opposed to relying upon human expertise.

While DNNs are often categorized within the broader spectrum of machine learning techniques, it is important to highlight a key distinction: traditional "shallow" ML approaches (\eg Random Forest) require manual selection of relevant features by humans to address specific problems. In contrast, DNNs eliminate this requirement by possessing the innate capability to autonomously identify and extract pertinent features necessary for solving the task at hand. This fundamental change in approach makes DNNs, and by extension, deep learning solutions, as more adaptable and superior options, particularly when dealing with large-scale datasets from which hidden patterns and underlying features must be learned.
In light of this however, a fundamental question arises: \emph{Is it possible for AI techniques that utilize deep learning methods to access a sufficient number of data points for effective learning?}

To address this question, we can review the explosion of software development data found on platforms such as GitHub, which, by January 2023, has accumulated over 100 million developers and  $\sim$400 million repositories  \footnote{\url{https://en.wikipedia.org/wiki/GitHub}}. Such a sheer amount of data has facilitated the adoption of DL methods to enhance and automate various SE tasks, including --but not limited to-- bug-fixing \cite{tufano2019,Chen,mesbah2019deepdelta,hata2018learning}, code summarization \cite{leclair2020improved,haque2020improved,lin2021improving} and code completion \cite{svyatkovskiy2020intellicode,ciniselli2021empirical,svyatkovskiy2019pythia,liu2020multi,izadi2022codefill,li2017code}

The rest of this section is designed to take the reader on a journey from the early methodologies to the current, advanced applications of AI in SE such as GitHub Copilot \cite{copilot} and ChatGPT \cite{chatgpt}.

\subsubsection{Once Upon a Time}
Before the widespread adoption of DNNs for automating tasks in SE, methods such as probabilistic modeling and SBSE technique were commonly employed. These foundational techniques focused on leveraging mathematical frameworks and algorithms to solve optimization problems, predict outcomes, and make decisions. 

For instance, Bayesian methods \cite{gelman1995bayesian} offers a robust framework for tackling diverse challenges in SE by integrating prior knowledge and employing probabilistic reasoning. This approach facilitates informed decision-making in situations characterized by uncertainty and has been effectively utilized to support and automate tasks related to SE. This includes areas such as software defect prediction \cite{arar2017feature,okutan2014software,sunil2018bayesian,jain2006software,dejaeger2012toward,pandey2018software,rahim2021software}, prediction of software quality \cite{neil2003software,wagner2009bayesian,moraga2008evaluating,radlinski2011conceptual,ziv1997constructing}, and various aspects of software testing \cite{king2018towards,zhao2015clustering,mirarab2007prioritization,han2010evaluation,ozekici2001bayesian}, maintenance \cite{ziv1997constructing,bibi2016bayesian,del2023bayesian,de2008software,bouktif2014predicting} and  software reliability \cite{bai2005bayesian,singh2001bayesian,yang2013bayesian}.

Similarly, SBSE applies the principles of search algorithms, such as genetic algorithms \cite{holland1992genetic}, hill climbing \cite{minsky1961steps} and ant colony \cite{ant} --among others-- to find near-optimal solutions to complex SE problems. In this field, a significant amount of research has been dedicated to areas such as test case generation and prioritization \cite{lin2019towards,ali2009systematic,gupta2021imp,gol2005meta,scalabrino2021adaptive,prak2020pot,gupta2017ts,gupta2022ts,jin2002genetic,evosuite,stall2021imp},  software refactoring \cite{sbr2008,mkaouer2016use,kebir2017,nas2020}, program repair \cite{le2011genprog}, and UI generation\cite{linares2015gemma}.

For example, Le Goues \etal \cite{le2011genprog} presented \emph{GenProg}, an automated technique based on genetic programming that is designed for automatically repairing defects in software systems. Leveraging principles from evolutionary computation, particularly the use of genetic operators such as mutation and crossover, this approach seeks to develop patches (\ie fixes) that maintain the original functionality of the program --while being evolved.

These early methods have laid the groundwork for more sophisticated methods built on advanced learning-based techniques such as DNNs that characterize contemporary SE. 

Transitioning from probabilistic approaches and machine learning techniques that necessitate manual feature extraction, to those grounded on learning-based methods (\eg DNN), marks a significant advancement in the field. Within this framework, researchers have been developing DL methods for the automation of various SE-related tasks such as code comprehension, code generation, code summarization, bug-fixing and program translation, among others.

To gain a deeper insight into the functioning of these techniques, we will concentrate on three seminal works that have played a critical role in the application of DL methods for the automation of SE tasks, including: (i) \faBug~Bug-Fixing Activities \cite{tufano2019}, (ii)  \faFileText~Source Code Summarization \cite{hu2018deep}, and (iii) \faCog~Code Completion \cite{svyatkovskiy2019pythia}

 Before delving into the details of these three distinct studies, we must anticipate that, due to space limitations, we are unable to cover the numerous recent applications of deep learning in automating a variety of SE tasks. Nevertheless, for a complete list of software engineering-related tasks, we direct readers to the systematic literature review conducted by Watson \etal \cite{watson2022systematic}.

\smallskip

\noindent \faBug~Tufano \etal \cite{tufano2019} explore the efficacy of a Neural Machine Translation (NMT) method \cite{Bahdanau2014NeuralMT} for automating the bug-fixing process. Specifically, their model incorporates two distinct Deep Neural Networks (DNNs), an encoder and a decoder, which are trained using a dataset of bug-fix pairs (BFPs). These BFPs consist of pairs of text strings, where the initial string (input) is a Java method containing a bug, and the subsequent string (target) is the same Java method after the bug has been fixed.

Unlike other research in the realm of automated bug fixing, the methodology proposed by Tufano \etal \cite{tufano2019} has undergone testing across a diverse array of bugs, instead of focusing solely on particular bug types or warnings (\eg only considering single-line bugs~\cite{Chen}, or targeting compilation failures as discussed in \cite{mesbah2019deepdelta}).
\smallskip
 
\noindent \faFileText~Hu \etal \cite{hu2018deep} propose \emph{DeepCom}, a DL-based technique designed with the purpose of automatically generating code comments for Java methods. In approaching the problem of \emph{source code summarization}, Hu \etal formulate the task at hand as a machine translation problem which translates source code written in a programming language to comments in technical natural language. 
\smallskip

\noindent \faCog~Svyatkovskiy \etal \cite{svyatkovskiy2019pythia} present \emph{Pythia}, a comprehensive approach for AI-assisted code completion. This method is designed to produce ordered suggestions of methods and API calls that developers can utilize while coding. Pythia was developed by training on code contexts derived from the Abstract Syntax Tree (AST) and has been incorporated into the \emph{Intellicode} plugin, making it accessible within the Visual Studio Code IDE \footnote{It is the publicly available IDE developed by Microsoft (\url{https://code.visualstudio.com})}.

\subsubsection{Recent Advancements in Automating Software Engineering Practices}

Lately, there has been a notable transformation in the domain of automated SE, marked by a growing preference for pre-trained models using the Transformer architecture \cite{transformer}, drifting away from conventional DL techniques that utilize other solutions, such as Long Short-Term Memory (LSTM) network \cite{hochreiter1997long}.

The introduction of the Transformer model has been pivotal in enhancing the automation of tasks within SE. Specifically, from an architectural standpoint, Transformer models present two key benefits compared to other advanced DL techniques used in automating SE tasks \cite{tufano2019,hu2018deep,haque2020improved,tufano2019icse,iyer:acl,Tufano:icsme2019,Watson:icse2020}: (i) they offer greater efficiency than Recurrent Neural Networks (RNNs) such as LSTMs, by enabling parallel computation of output layers, and (ii) they excel at identifying both hidden and long-range dependencies between tokens (\ie sub-unit pieces of a string), without the assumption that tokens in closer proximity are more strongly related than those farther apart. This latter feature is particularly valuable for SE activities, especially for tasks involving code where --for example, a variable declaration is several code tokens apart from its first usage.

In addition, while conventional DL methods used for automating SE tasks typically adhere to a direct training approach—gathering a sufficient dataset and then training the model to identify task-specific features—the training process for a Transformer model unfolds in two distinct phases. Initially, the model undergoes pre-training on a large-scale dataset via self-supervised learning \cite{Delvin:2019}. This stage equips the model with a foundational understanding of the relevant language(s), using, for instance, input sentences in English and Spanish with a certain percentage (\eg 15\%) of their tokens being masked and prompting the model to predict these masked tokens \cite{Delvin:2019}. This self-supervised pre-training aims to instill within the model a broad linguistic comprehension, setting a solid base for its subsequent task-specific training (\eg Sentence Translation).
The next step involves fine-tuning (\ie specializing) the model through supervised learning (\eg by training on pairs of English sentences and their Spanish translations). This phase precisely tailors the model for the designated task.

The ability to equip the model with foundational knowledge through pre-training, sets the stage for more effective fine-tuning. In addition, a further significant advantage of using pre-trained DL-based solution is the capacity these model posses in applying the insights gained during the initial training phase (\ie pre-training) to subsequent tasks, even in situations where there is a shortage of data for fine-tuning. In this direction, the work by Robbes and Janes \cite{robbes2019leveraging} underscores the importance of pre-training in tackling the challenges posed by small datasets in SE tasks, with a particular focus on \emph{sentiment analysis}, by transferring knowledge from the pre-training stage to specific downstream tasks.

These benefits, combined with the breakthrough Transformer architecture, address several shortcomings of earlier DL methods, such as RNNs, which were previously utilized by developers for automating SE practices. As a result, this advancement has facilitated the automation of tasks that were once considered extremely challenging due to existing limitations. 

Notably, innovations have emerged in various domains, including the automation of logging activities \cite{mastropaolo2022lance}, the completion of GitHub Workflow files \cite{mastropaolo2023automatically}, improvements in code review processes \cite{tufano2022using,li2022automating,li2022codereviewer}, and advancements in testing automation \cite{tufano2022generating,tufano2020unit}, among others \cite{zhang2023pre,wang2022no,mastropaolo2024towards,liu2023ccrep,wang2023codet5}. This progression signifies a remarkable leap forward in the automation capabilities within SE.


Moreover, recent developments in DL have led to the rise of DNNs with billions/trillions of parameters \cite{gemini}, marking significant milestones in AI research. Technologies like ChatGPT \cite{chatgpt} and Gemini \cite{gemini} exemplify this trend, with their extensive pre-training on vast text corpora enabling them to produce contextually relevant and coherent text across diverse subjects. The sheer size of these models and their comprehensive pre-training underpin their extraordinary capabilities for generating and understanding  text , far beyond what traditional DL models featuring up to million of parameters can achieve. 
This trend, which reveals the emergence of a new class of DL-based solution anchored in Large Language Models (LLMs) and Foundation Models (FMs) is playing a pivotal role in different fields, with SE being no exception.

To explore the distinctions between methodologies based on LLMs/FMs versus ``smaller'' DL models --for automating SE practices-- we review the research by Mastropaolo et al. \cite{mastropaolo2021studying}. Their work utilized pre-trained Transformer models, specifically a Text-to-Text Transfer Transformer (T5) model \cite{raffel2020exploring} with  $\sim$60 million parameters, pre-trained on a mix of code and technical natural English (\eg, code comments). This pre-training allowed the transfer of learned knowledge to tasks like bug fixing, code summarization, code mutation, and testing, setting new benchmarks in these areas.

Conversely, the training of methods such as ChatGPT although grounded upon the foundation approach of ``pretrain-then-finetune'', it extended in a significant way. For example, ChatGPT, through its training on a diverse range of internet text, not only learns from a broad dataset during pre-training but also leverages reinforcement learning from human feedback (RLHF)  \cite{christiano2017deep} to refine its responses to be more aligned with human preferences. This process does not strictly fall under the traditional ``fine-tuning'' phase but represents an additional layer of model refinement and customization.  


The principal distinction between conventional pretrain-then-finetune models and Foundation Models (FM), like ChatGPT or tools such as GitHub Copilot \cite{copilot}, centers on the magnitude of pre-training, the specifics and breadth of the fine-tuning process, and the integration of supplementary training techniques such as RLHF. These variances equip LLMs and targeted applications with the capability to excel in a broad array of tasks, even those not explicitly included in their initial training datasets. They can also adjust in real-time to the specific inputs of users within their respective application areas, by responding to specialized prompts that users provide the knowledge acquired by the model during its extensive pre-training phase.

In light of this, it's important to note that solutions like ChatGPT and GitHub Copilot have garnered significant attention from researchers in the domain. This interest is driven by the desire to explore the capabilities of FM and LLMs in supporting the automation of tasks related to SE. Additionally, there is a keen interest in identifying any potential risks or pitfalls associated with these models that might temper the current enthusiasm shared by both researchers and practitioners. The aim is to gauge the extent to which these advanced models can contribute to the field \cite{sobania2021choose,nguyen2022empirical,vaithilingam2022expectation,liu2024your,white2023chatgpt,sobania2023analysis,ahmed2024automatic,tufano2024unveiling} while being mindful of any challenges or issues that may arise from their use \cite{pearce2021empirical,howard2021github,imai2022github,liu2023refining,mastropaolo2023robustness,yang2024robustness,ahmed2024studying}.

\vspace{-0.3cm}
\section{Is it the fall of Software Engineers?}
\label{sec:discussion}

The advent of LLMs and FMs marks a pivotal shift in software development, transforming coding practices, and problem-solving. These AI-driven tools, by supporting and enhancing SE practices, promise significant productivity boosts, fresh problem-solving techniques, and a possible redefinition of SE roles. They pave the way for more efficient automation, fostering collaborative efforts between humans and AI. However, there is no free-lunch, and embracing these technologies also requires addressing ethical dilemmas and challenges to ensure responsible utilization and integration into the industry.
This section explores cutting-edge AI technologies in SE, starting with an overview of the current AI-driven solutions landscape. We will then scrutinize both the advantages and challenges associated with these technologies. Concluding, we reflect on how the integration of AI into development processes is transforming the role of software engineers, focusing on the shift in practices and skill sets that are shaping the future of SE.%

\vspace{-0.5cm}
\subsection{Emerging technologies and capabilities of AI-driven systems for SE and beyond}

As researchers and practitioners delve deeper into the capabilities of Large LLMs and FMs, we are witnessing significant breakthroughs in different fields, including SE. These models not only impact SE practices such as code generation and summarization but also promise to revolutionize the operational workflow, making the development process more efficient and innovative.
For example, Cognition Labs recently introduced Devin \cite{devin}, an AI that pioneers the role of a software engineer. This AI's capabilities extend to planning, task execution, and the full cycle of software development, including autonomous bug fixes and feature implementations in open-source projects. Moreover, Devin stands out by being able to train its own AI models, showcasing an advanced level of autonomy in SE tasks never seen before. Another example of AI, not specific for SE, is AutoGPT \cite{autogpt} which allows the automation of several tasks with little human intervention; for instance tasks such as pipelines creation and execution can be easily done with AutoGPT.

Within this framework, other AI-driven solutions which are worth discussing are \emph{Claude2} \cite{claude2}, \emph{Gemini} \cite{gemini} and \emph{Sora} \cite{sora}.  Claude2 \cite{claude2} is an advanced AI system proficient in understanding and generating text that resembles human writing, achieving around 72\% on the HumanEval dataset \cite{chen2021evaluating} when tested on code-related tasks. Its applications however, span across various fields, not just SE. Remarkably, Claude2 scored 76.5\% on the BAR exam's multiple-choice section, demonstrating its ability to process and analyze extensive information, thanks to its capability of handling up to 100K tokens in a single prompt. This feature enables Claude2 to digest and interact with vast amounts of text effectively, even books.

In December 2023, Google presented \emph{Gemini} \cite{gemini}. Boasting 175 trillion parameters in its most powerful configuration, Gemini is the biggest and most advanced AI model ever created. It has revolutionized the idea of AI-powered tools for human assistance.  While it excels in automating SE tasks \cite{geminiteam2023gemini}, what truly sets it apart is its ability to handle various data formats, including images, videos, audio, and text opening up entirely new possibilities for applications in fields like scientific discovery, creative content generation, and even real-time language translation. 
Within the domain of SE, Gemini offers considerable help for a variety of purposes. Its capabilities in code generation have demonstrated superior performance compared to the former frontrunner, ChatGPT \cite{chatgpt}. This advancement opens up several potential avenues for further investigations, particularly in view of recent studies that highlight the potential of LLMs to repair software by being used as code generators.

Building on the advancements in AI-powered systems, OpenAI introduced Sora \cite{sora}. This groundbreaking text-to-video model lets users create realistic and imaginative scenes by providing a textual description as input. While not yet publicly  accessible, it is conceivable to predict that SE researchers and practitioners will leverage the latest creation by OpenAI \ie Sora, for generating tutorials and interactive content, such as videos, for various applications.

The potential of AI-powered systems such Devin, AutoGPT, Claude2, Gemini, and Sora is undeniable, nonetheless, we must acknowledge that this rapid advancement comes with its own challenges and drawbacks. In the upcoming sections, we will explore essential aspects for researchers and practitioners as they integrate these advanced technologies into their work. This includes a thorough examination of the critical factors that must be taken into account with the adoption and use of these powerful tools across fields and SE in particular.
 
 \vspace{-0.5cm}
\subsection{The Double-Edged Sword: Threats of Emerging AI Technologies}

The rapid rise of AI-powered systems presents a double-edged sword. On the one hand, they offer immense potential to revolutionize various fields, including SE. On the other, we are now at the beginning of a new set of challenges and threats associated with the use of these powerful tools. While these threats include society, professionals and ethical elements, the field of SE faces some unique challenges. From a broader perspective, as AI systems automate tasks across various industries, job displacement is a major concern \cite{frey2017future}. In particular, while new opportunities can emerge, the transition could be disruptive for many workers whose job mainly revolves around repetitive tasks for which AI-driven solution can provide better support and full automation.  

AI systems trained on extensive (and public) datasets may reflect and amplify existing societal biases, known as ``algorithmic bias'' affecting decisions in areas like finance, employment, and law enforcement. Additionally, the opaque workings of some AI models challenge transparency, leading to concerns over accountability and fairness in high-stakes decisions. Beyond those, security vulnerabilities are other threats hidden behind the use of these technologies as AI-system themselves can be vulnerable to hacking activities and malicious manipulation. For instance, there are different types of "poisoned models" \cite{Minghong2020,kurita2020weight} that could be a threat for the adequacy of AI-enabled systems because bad data is injected during models training to skew their outputs. 
Furthermore, as models behind AI tools become larger, with increasing parameters, essential questions about energy efficiency and carbon emissions come to the forefront~\cite{de2023growing,wu2022sustainable,chien2023reducing,trihinas2022towards}.
When we look at the professional landscape of SE, we soon realize that beside the above mentioned challenges, there are unique ones that impact the field and for which researchers and practitioners should take adequate measure. For example, in a recent research Dakhel \etal \cite{dakhel2023github} revealed the dual nature of GitHub Copilot, meaning that Copilot quickly becomes a liability for novice and beginners, while being an asset for experts, enhancing their productivity and efficiency. In addition, Imai \cite{imai2022github} delved into the extent to which Copilot serves as a viable substitute for a human pair programmer. Their findings revealed that while Copilot led to enhanced productivity (\ie an increase in the number of added lines of code), it also resulted in diminished quality of the generated code.
In a different research, Pearce \etal \cite{pearce2022asleep} aimed at investigating whether Copilot's generated code was affected by any vulnerability issue. The results highlighted for $\sim$40\% of the tested elements, the code automatically recommend by the state-of-the-art code generators presented vulnerabilities. In this regard, also Mastropaolo \etal \cite{mastropaolo2023robustness} showcased how the code synthesized by GitHub Copilot varies if the model is prompted with different but, semantically equivalent natural language descriptions informing the model on the behaviour of the code that it has to be generated.

Moving along these lines, Tufano \etal \cite{tufano2024unveiling} undertook an empirical investigation into how ChatGPT is used to automate tasks related to SE practices. The result of their study was a taxonomy accompanied by actionable advice for practitioners and researchers on leveraging ChatGPT for automating SE tasks.

Other researchers broadly focus on properties pertaining AI-driven techniques for SE automation. In a recent investigation, Yang \etal \cite{yang2024robustness},  performed the first Systematic Literature Review (SLR) on AI-enabled systems within SE. 
Their thorough analysis identified six essential non-functional requirements: (i) Robustness; (ii) Security; (iii) Privacy; (iv) Explainability; (v) Efficiency; and (vi) Usability, pertaining LLM4Code adopted in SE automation practices. The broad spectrum of issues that could emerge with the utilization of AI-driven system (\eg ChatGPT) for automating SE-related practices, also includes critical elements such as \emph{legal compliance}, \emph{ethical governance}, \emph{trust}, \emph{accountability}, \emph{transparency}  \emph{fairness} and \emph{intellectual property} being just the tip of the iceberg. For instance, when a company allows employees to use AI systems, without clear guidelines, the underlying risk is the erosion of trust within software development teams and even the possibility of negatively impacting critical infrastructures. The responsibility for errors or biases introduced by AI, such as in code or documentation, also becomes a concern within this operational framework. 
Additionally, the long-term effects of integrating AI into SE practices need consideration, especially regarding its impact on future engineers.
This analysis prompts a reflection on the evolving role of software engineers in an era increasingly defined by AI, highlighting the challenges that distinguish the new generation from its predecessors. Thus, in \secref{sub:human_element}, we introduce tangible opportunities for software engineers (\faLightbulbO) as well as the potential challenges they may face (\faWarning) when it comes to the use of AI-driven system to aid in software development.

\vspace{-0.5cm}
\subsection{The Human Element: Navigating the Software Engineering of The Future}
\label{sub:human_element}

\subsubsection{Software engineers as datasets creators and validators of AI-based generated code:}

Software engineers might become responsible for creating and validating the vast datasets used to train large language models for task automation, as well as verifying and validating the quality of AI-based systems and AI-generated code. Model mismatch is a phenomenon that require more attention from out community \cite{lewis2021}.

\faWarning~This raises concerns about potential biases present in the data, which could be amplified by the LLM and lead to discriminatory or flawed code generation. Additionally, the sheer volume of data required for effective LLM training creates a significant workload potentially redirecting them from core development tasks.

\faLightbulbO~However, this is a tremendous opportunity for creating quality models (for AI-systems) as we have done in the SE community, as well as transferring to the AI community our learned lessons in developing code for large systems. There are already some initial efforts  on adopting SE for AI-based systems development \cite{Amershi2019,Serban2020}.

\vspace{-.2cm}
\subsubsection{Should humans generate production/testing code?:}  While AI can generate code, the question of when to use human-written code versus AI-generated code becomes crucial. Production code, which forms the critical skeleton of software systems, requires a deep understanding of system architecture, security considerations, and long-term maintainability. Similarly, for crucial testing scenarios, human expertise in crafting comprehensive test cases might be irreplaceable.

\faLightbulbO~It is safe to state that human expertise in these areas is still crucial, as AI-generated code could introduce vulnerabilities and increase long-term maintenance costs, thus being detrimental for the overall software development pipeline. Therefore, code generation and automation capabilities of AI agents could be used as a way for rapid prototyping and Minimum Viable Products generation.

\vspace{-.2cm}
\subsubsection{Interpretability/Understandability/ of AI-systems generated artifacts:}  
Many AI models, especially LLMs, can be opaque in their decision-making processes.  This lack of interpretability makes it difficult for software engineers to understand how AI-generated artifacts have been produced.

\faLightbulbO~Debugging and maintaining such code becomes challenging, potentially leading to hidden vulnerabilities or unexpected behavior in the software. Nonetheless, with human-crafted test cases for this purpose, the issue can be mitigated.  Moreover, readability of AI-generated code is an interesting avenue for research.

\vspace{-.2cm}
\subsubsection{Context and code generation for large software systems: } AI code generation tools might not fully grasp the intricate context and complexity of large software systems. 

\faLightbulbO~For optimal code production, AI must understand the system's purpose, its existing code, and future objectives comprehensively. Moreover, within vast systems where interconnected elements proliferate, grasping the foundational components crucially depends on human expertise, highlighting areas where AI's current capabilities may fall short.

\vspace{-.2cm}
\subsubsection{Can AI models have project management capabilities?: }
Project management encompasses strategic planning, the distribution of resources, evaluating risks, and communication—domains that require human assessment, expertise, and flexibility.

\faLightbulbO~Although AI can aid in organizing tasks and analyzing data related to project management, the intricate decisions related to guiding the project's course and managing team interactions are still beyond the reach of AI's abilities. Anyway, AI agents to support risks evaluation and mitigation based on historical data are an interesting avenue for research.

\vspace{-.2cm}
\subsubsection{Intellectual Property (IP) of AI-generated artifacts: } The question of who owns the intellectual property rights associated with AI-generated artifacts remains a legal grey area.

\faWarning~As software engineers provide training data and potentially refine AI-generated outputs, the lines of authorship become blurred. Clear legal frameworks will be crucial to address IP ownership and copyright issues in AI-assisted software development.

\vspace{-.2cm}
\subsubsection{Problem-solving/logic/reasoning capabilities of humans and AI systems: } AI excels at handling repetitive tasks and identifying patterns, yet it falls short of the human capacity for inventive problem-solving, sophisticated logic, and detailed reasoning. 

\faWarning~The field of SE frequently demands confronting distinctive problems and crafting original solutions. Excessive dependence on AI might negate the development of essential critical thinking abilities among upcoming students and software engineers. However, AI could be a supporting tool for envisioning different scenarios as in the case of defining software architectures or generating test cases for complex systems. 

\vspace{-.2cm}
\subsubsection{AI systems as tutor for software engineers: }

AI agents could serve as additional educational resources for software engineers, yet exclusive reliance on AI for software engineering training presents constraints. 

\faLightbulbO~ Human teachers deliver vital guidance, contextual knowledge, and the capacity to address complex inquiries that might surpass the present abilities of AI technologies. Nonetheless, AI agents have the potential to customize learning trajectories and offer specific practice scenarios, thereby playing an instrumental role as mediator in the field of SE education.

\section{Conclusion: What is next?}
\label{sec:conclusion}

In the evolving landscape of software development, the integration of AI systems, presents a complex matrix of opportunities and challenges. Software engineers are increasingly becoming integral to the development, deployment, and maintenance of AI systems. As these technologies continue to advance, the role of software engineers is not only expanding but also diversifying.  They are now required to possess a deep understanding of those critical aspects underpinning these technologies, such as machine learning algorithms, data management, and ethical considerations, in addition to traditional SE skills like coding proficiency and system design. We must be aware that software engineers have adopted ML methods as components-of-the-shelf and black boxes, which has led to challenges, pitfalls, and mistakes when using ML~\cite{mojicahanke2023machine}, therefore it is a paramount to educate software engineers on this matter and provide them with supporting tools  \cite{mojicahanke2023machine,cabra2023}.
This transition is not likely to happen overnight, and thus proper training of the future engineers of the future is imperative. Educational institutions, universities and organizations must adapt their curricula and training programs to include elements of artificial intelligence, alongside the main pillar of the selected curricula. This holistic approach to education will equip future software engineers with the skills needed to navigate the complexities of AI integration. 

Software engineers currently face an unprecedented level of responsibility, as their roles are set to evolve and integrate into a novel environment dominated by AI-driven systems. These advanced systems are poised to direct the future of software systems and infrastructure, forming the backbone of our society, and thus we all are better being prepared to embrace and navigate this transformative era with agility and foresight.

\balance
\bibliographystyle{ACM-Reference-Format}
\bibliography{main}

\end{document}